\documentclass[9pt,twocolumn,twoside]{opticajnl}
\journal{opticajournal} 

\setboolean{shortarticle}{false}

\usepackage{lineno}
\usepackage{graphicx}
\usepackage{tensor}
\usepackage{xcolor}

\title{Stokes and skyrmion tensors and their application to structured light}

\author[1]{Stephen M. Barnett}
\author[1]{Sonja Franke-Arnold}
\author[1,*]{Fiona C. Speirits}

\affil[1]{School of Physics and Astronomy, University of Glasgow, Glasgow G12 8QQ, UK}

\affil[*]{fiona.speirits@glasgow.ac.uk}

\begin{abstract}
We replace the familiar Stokes vector by a tensor.  This allows us to introduce, for example, polar-coordinate components of the Stokes vector. From the tensor we can 
derive the skyrmion field for mapping the polarization in structured light beams.  
These ideas have wider application in optics and in electromagnetic theory.  We
illustrate this with an example from non-paraxial optics and for Poynting's
vector.
\end{abstract}

\setboolean{displaycopyright}{false} 

\begin{document}

\maketitle

\section{Introduction}\label{sec_1}
The study of structured light has introduced us to a remarkable variety of topological features.  Some of these have been engineered for specific purposes \cite{Road}, while
others exist in naturally occurring optics \cite{Nye}.  Important examples include light with phase singularities, often associated with the orbital angular momentum
of light \cite{Les,OAMBook,PhilTrans} and features associated with the spatial 
variation of the local orientation of the electric field vector \cite{Nye,NyeHajnal,BerryDennis}. The latter includes light beams with radial or 
azimuthal polarization, which exhibit remarkable focusing properties \cite{Quabis}.

Here we investigate the features of polarization patterns and other features of the structured electromagnetic field by means of the skyrmion field, familiar from the
physics of magnetic structures \cite{Seki,Evershor,Goebel,Tokura,Petrovic}.
To do so, we generalize the familiar Stokes parameters by the introduction of the
tensor formalism.  This enables us to investigate structured light in coordinate 
systems other than the familiar $x$, $y$ and $z$ Cartesian coordinates.  As is 
often the case in physics, the associated ability to use a coordinate system
reflecting the symmetry of a given problem can simplify analyses and suggest new
physical insights.  A simple example is the vortex beam pattern produced by a 
glass cone \cite{Cone}.  Replacing the Stokes vector by a tensor has the 
additional benefit of introducing the skyrmion tensor, the field lines of which 
map out lines of constant polarization \cite{Joerg}.

The tensor formalism can readily be applied to other vector quantities, including those 
associated with non-paraxial light.  We demonstrate this by applying it to quantities formally associated with vector singularities, the L-lines and C-lines, for 
arbitrary (monochromatic) light fields \cite{Nye,NyeHajnal,BerryDennis}.  The light 
field radiated by a single dipole provides a simple but informative example.  Finally,
the skyrmion tensor can be applied to a wide variety of vector fields beyond
optical polarization.  We illustrate this by reference to the Poynting vector for 
a radiating dipole and the Newtonian gravitational field of a point mass.

\section{Preliminary details}

Our work brings together three somewhat disparate ideas: Stokes parameters, skyrmions
and skyrmion fields, and the calculus of tensors.  For this reason we present, here, brief introductions or reminders on each of these.  The more experienced reader may 
wish to skip to section 3.

\subsection{Stokes parameters}\label{sec_2a}

Stokes parameters have long been employed to map out the polarization in an optical
field \cite{Born,Wolf,Hecht}.  Without exception, to the best of our knowledge, these have 
always been defined in terms of three Cartesian components: $s_3$, which is usually 
associated with the difference in the intensities of right and left circular 
polarization, while $s_1$ and $s_2$ are associated with the differences between pairs 
of orthogonal linear polarization.  The set is completed by $s_0$, which is the total 
intensity.

We shall be interested in the local state of polarization irrespective of the 
intensity and so work with the normalized Stokes parameters, $S_i = s_i/s_0$.  These
three parameters may conveniently be mapped onto the surface of a sphere, the 
Poincar\'{e} sphere \cite{Born}.  This sphere sits in an abstract space, but we can 
associate it with our familiar three-dimensional space by mapping the $1$, $2$ and $3$ 
directions onto Cartesian axes: $1 \rightarrow x$, 
$2 \rightarrow y$ and $3 \rightarrow z$.  If we do this then it becomes natural to ask 
if we can use other coordinate systems, such as cylindrical or spherical polar 
coordinates and, if we can, what the non-Cartesian components of the Stokes vector can 
tell us about optical polarization patterns.

We address this question by treating the Stokes vector as a rank-one tensor.  This
allows us to map out variations in the polarization associated with structured light 
fields without being restricted to Cartesian coordinates.  We shall find, for 
example, that polarization patterns created by superposing orthogonally polarized
Laguerre-Gaussian modes \cite{Siegman,Eberly} may be analyzed, naturally, using
Stokes tensors expressed in cylindrical polar coordinates.

\subsection{Skyrmions}

Skyrmions were famously introduced by Skyrme in the study of non-linear field
theory, with application to nuclear physics~\cite{Skyrme1,Skyrme2}.  The idea
has had greatest impact, however, mostly in the guise of so-called baby
skyrmions, in other fields of physics
\cite{Piette,Vollhardt,Volovik,Leggett,Mechelen,Wilczek,Pisamen, Vilenkin}. It
is probably in magnetic media where the concept has demonstrated the greatest
potential, certainly for practical applications
\cite{Seki,Evershor,Goebel,Tokura,Petrovic,Sachdev,Bogdanov,Rossler,Muhlbauer,Romming,Foster}.

The application of Skyrme's ideas to optics is relatively new, but already
skyrmions and skyrmionic structures have been identified in a wide variety of
contexts~\cite{Tsesses,Shi,Shen2021a,Shen2021b,Cuevas,Shen2024,Skuse}. It is in the
field of paraxial optics where optical skyrmions have, perhaps, their simplest
manifestations, associated with polarization patterns in structured light
\cite{Gao,Zhu,McWilliam}, and offer the largest degree of freedom in exploring different geometries.  This may be due to the comparative ease with which optical skyrmions can be
prepared in the laboratory. We have shown that, in this paraxial regime, the
skyrmion field lines are simply lines of constant polarization~\cite{Joerg}. A
comprehensive theoretical study of paraxial optical skyrmions can be found in
\cite{Zhujun}.  In their simplest form, paraxial optical skyrmions arise from a
skyrmion field defined to be
\begin{equation}
\label{Eq.1}
\Sigma_i = \frac{1}{2}\varepsilon_{ijk}\varepsilon_{\ell mn}S_\ell\frac{\partial S_m}{\partial x_j}\frac{\partial S_n}{\partial x_k} ,
\end{equation}
where we have employed the summation convention, with an implied summation for
repeated indices over the three Cartesian coordinates, $x$, $y$ and $z$.  The
$\varepsilon_{ijk}$ and $\varepsilon_{mnp}$ are the familiar Levi-Civita
alternating symbols and $S_\ell$, $S_m$ and $S_n$ are the normalized (Cartesian)
components of the Stokes vector.

It is important to realize that the formula for the skyrmion field given above
applies, strictly, to Cartesian coordinates only. We cannot use it, for
example, to work with cylindrical polar coordinates to obtain $\Sigma_\rho$,
$\Sigma_\phi$ and $\Sigma_z$ from $S_\rho$, $S_\phi$ and $S_z$.  This is
regrettable, as the simplest optical skyrmions have cylindrical symmetry
\cite{Gao} and it would seem natural to employ cylindrical polar coordinates to
study them.  We address this short-coming by replacing the Stokes and skyrmion
vectors by rank-one contravariant tensors and employing covariant derivatives
in place of the partial derivatives appearing in eqn.~(\ref{Eq.1}).

\subsection{Tensors}

Tensors, as opposed to the more familiar vector fields, were introduced into
physics primarily for two reasons: the first is that they allow for the
formulation of a simple coordinate-independent scalar product in the form of a
scalar quantity; the second (and more important) reason is that they make
possible the application of differential calculus in a coordinate-independent
manner.  It is for this reason that special relativity and, more importantly,
general relativity are formulated in terms of tensor quantities
\cite{Weinberg,Dirac,Schutz}.  Tensors have a general applicability outside
of relativity, however, and they are frequently employed in diverse areas 
of the physical sciences \cite{Jeffreys,Goodbody,KayTensors,Stephenson,Frankel}.

The first thing to recall about tensors is that there are fundamentally two
distinct types: the contravariant tensor (with raised indices) and the
covariant tensor (lowered indices). The difference between
these lies in the manner in which they transform under coordinate
transformations. If we transform from the coordinates $x^i$ to $x'^j$ then the
two types of tensor transform in different ways:
\begin{align}
\label{Eq.2}
A'^j &= \frac{\partial x'^j}{\partial x^i}A^i \nonumber \\
A'_k &= \frac{\partial x^\ell}{\partial x'^k}A_\ell .
\end{align}
In each of these the index $\ell$ appears twice and so there is an implied 
summation over the three (general) coordinates.  The forms of the two types
of tensor ensures that the scalar product (formed from a contravariant and a
covariant tensor) is, indeed, coordinate independent:
\begin{align}
\label{Eq.3}
A'^jA'_j &= A^i\frac{\partial x'^j}{\partial x^i}\frac{\partial x^\ell}{\partial x'^j}A_\ell  \nonumber \\
&= A^iA_\ell \delta^\ell_i  \nonumber \\
&= A^iA_i .
\end{align}
It also means that, in contrast to vector components as we usually encounter
them in optics, the components of $A'^j$ and $A'_\ell$ may have different
dimensions and also different dimensions to the Cartesian components.  For
example, the azimuthal components of the contravariant and covariant tensors in
cylindrical polar coordinates are
\begin{align}
\label{Eq.4}
A^\phi &= \frac{\partial \phi}{\partial x}A^x + \frac{\partial \phi}{\partial y}A^y  \nonumber \\
&= -\frac{\sin \phi}{\rho} A^x + \frac{\cos \phi}{\rho} A^y ,  \nonumber \\
A_\phi &= \frac{\partial x}{\partial \phi}A_x + \frac{\partial y}{\partial \phi}  \nonumber \\
&= -\rho\sin\phi A_x  + \rho\cos\phi A_y .
\end{align}
Clearly the dimensions of these two quantities differ by the square of a distance.

The description of the tensor quantities is completed by replacing ordinary
partial derivatives by covariant derivatives.  Before presenting these we need
to take a small digression to introduce the metric tensor $g_{ij}$, the form of
which encodes the requisite properties of the coordinate system of choice.  It
appears, naturally, in a generalization of the infinitesimal form of
Pythagoras's theorem:
\begin{equation}
\label{Eq.5}
ds^2 = g_{ij}dx^idx^j  .
\end{equation}
For Cartesian coordinates this reduces to the familiar form
\begin{equation}
\label{Eq.6}
ds^2 = dx^2 + dy^2 + dz^2 \quad \Rightarrow g_{ij} = \delta_{ij} .
\end{equation}
For cylindrical and spherical polar coordinates, however, the metric tensor is position
dependent:
\begin{align}
\label{Eq.7}
ds^2 = d\rho^2 + \rho^2 d\phi^2  + dz^2
&\Rightarrow \ g_{\rho\rho} = 1, \: g_{\phi\phi} = \rho^2,
\: g_{zz} = 1 ,  \nonumber \\
ds^2 = dr^2 + r^2\sin^2\theta d\phi^2 + r^2 d\theta^2 &\Rightarrow g_{rr} = 1, \:
g_{\phi\phi} = r^2\sin^2\theta , \: g_{\theta\theta} = r^2 .
\end{align}
(Here repeated indices over specific coordinates $\rho$, $\phi$, $z$, $r$,
$\phi$, $\theta$ are not summed.)  We require, also, the contravariant form of
the metric tensor that gives $g^{ij}g_{jk} = \delta^i_k$, which we can think of
as the identity matrix, so  the contravariant form of the metric is simply the
matrix inverse of the covariant form. The forms of the covariant derivatives,
which we denote by a subscript semicolon, is different for the contravariant
and the covariant derivatives.  The partial derivatives are replaced by
\begin{align}
\label{Eq.8}
\frac{\partial A^i}{\partial x^j} \rightarrow \tensor{A}{^i_;_j} &= \frac{\partial A^i}{\partial x^j} + \tensor{\Gamma}{^i_j_k}A^k
\nonumber \\
\frac{\partial A_i}{\partial x^j}  \rightarrow \tensor{A}{_i_;_j} &= \frac{\partial A_i}{\partial x^j} - \tensor{\Gamma}{^m_{i}_j} A_m ,
\end{align}
where $\tensor{\Gamma}{^i_j_k}$ is the affine connection, or simply the
connection, defined to be
\begin{equation}
\label{Eq.9}
\tensor{\Gamma}{^i_j_k} = \frac{1}{2}g^{im}\left(\frac{\partial g_{jm}}{dx^k} + \frac{\partial g_{km}}{\partial x^j} - \frac{\partial g_{jk}}{\partial x^m}\right) .
\end{equation}
Note that this is symmetric in the lower two indices: $\tensor{\Gamma}{^i_j_k}=\tensor{\Gamma}{^i_k_j}$. For the Cartesian coordinates the connections are zero and the covariant
derivatives are equivalent to the partial derivatives.  For other coordinate
systems, however, this is not true. We now have all of the tools necessary to
replace the Stokes parameters and skyrmion field by tensors.

\section{Stokes and skyrmion tensors}\label{sec_3}

The (normalized) Stokes parameters are formulated to correspond
to the three Cartesian components of the Poincar\'{e} sphere and a combination
of these provides a complete description of the optical polarization
\cite{Born,Hecht}.  For a structured beam these parameters depend on the position within the beam profile
and so we can conceive of local Stokes parameters as forming a vector field and
from this construct the corresponding skyrmion field.  It is important to
appreciate that this Stokes field vector is oriented in an abstract space and
it does not correspond to any direction in physical space.  We can map it onto
physical space, as stated in section \ref{sec_2a}, simply by mapping the numbered components $S_1$, $S_2$
and $S_3$ onto spatial directions: $S_1 \rightarrow S_x$, $S_2 \rightarrow S_y$
and $S_3 \rightarrow S_z$\footnote{It is certainly possible to choose other mappings. These would correspond, for example, to placing horizontal and vertical linear polarizations at the poles of the Poincar\'{e} sphere rather than the usual circular polarizations. The form of the skyrmion field is independent of that choice~\cite{Zhujun}.}. With this mapping implemented, we can evaluate
the spatial Cartesian components of the skyrmion field.  For example
\begin{align}
\label{Eq.10}
\Sigma_z &= S_z\left(\frac{\partial S_x}{\partial x}\frac{\partial S_y}{\partial y} -
\frac{\partial S_x}{\partial y}\frac{\partial S_y}{\partial x}\right) 
+ S_x\left(\frac{\partial S_y}{\partial x}\frac{\partial S_z}{\partial y} -
\frac{\partial S_y}{\partial y}\frac{\partial S_z}{\partial x}\right)
\nonumber \\
& \qquad + S_y\left(\frac{\partial S_z}{\partial x}\frac{\partial S_x}{\partial y} -
\frac{\partial S_z}{\partial y}\frac{\partial S_x}{\partial x}\right) ,
\end{align}
for which the skyrmion number for a paraxial beam propagating in the $+z$ direction is
\begin{equation}
\label{Eq.11}
n = \frac{1}{4\pi}\int\int dx dy \: \Sigma_z .
\end{equation}

As has been noted above, the simplest paraxial skyrmions have cylindrical
symmetry and this rather suggests that simplifications, or at least new
insights, might arise if we could formulate the Stokes vector and the skyrmion
field as tensor quantities and thereby employ cylindrical polar coordinates,
for which the components of the contravariant Stokes tensor are 
\begin{align}
S^\rho &= \cos\phi S^x + \sin\phi S^y  \nonumber \\
S^\phi &= \frac{1}{\rho}(-\sin\phi S^x + \cos\phi S^y) \nonumber \\
S^z &= S^z .
\end{align}
As a further illustration, the components of the Stokes tensor in spherical polar coordinates are
\begin{align}
S^r &= \sin\theta\cos\phi S^x + \sin\theta\sin\phi S^y + \cos\theta S^z  \nonumber \\
S^\theta &= \frac{1}{r}(\cos\theta\cos\phi S^x + \cos\theta\sin\phi S^y - \sin\theta S^z) \nonumber \\
S^\phi &= \frac{1}{r\sin\theta}(-\sin\phi S^x + \cos\phi S^y) . 
\end{align}
An additional benefit may occur when describing the focusing of structured light
beams or, more generally, a cylindrically symmetric operation of any object on
the beam~\cite{Cone}. Here we would expect the correctly formulated azimuthal
component of the Stokes tensor to be essentially unchanged by the process.
Moreover, as we shall see, this component of the Stokes tensor can be zero for
the simplest skyrmions, leaving only $S^\rho$ and $S^z$ to contribute.

The great benefit of the tensorial formulation is that by
constructing a desired quantity (in our case the skyrmion field) from
established tensor quantities (the Stokes tensor) and using only covariant
derivatives, we guarantee that the desired quantity is also a tensor, with all
the corresponding advantages. It follows, then, that the skyrmion
(contravariant) tensor field is
\begin{equation}
\label{Eq.12}
\Sigma^i = \frac{1}{2}\varepsilon^{ijk}\varepsilon_{\ell mn}S^\ell \tensor{S}{^m_;_j}\tensor{S}{^n_;_k} .
\end{equation}
The skyrmion tensor retains the most important properties of the skyrmion
vector. Among these we have that it is transverse as
\begin{equation}
\label{Eq.13}
\tensor{\Sigma}{^i_;_i} = 0 .
\end{equation}
This follows directly from the fact that $\tensor{\Sigma}{^i_;_i}$ is a scalar
and hence independent of the coordinate system used.  We know that it is zero
when evaluated in Cartesian coordinates and it necessarily follows that it is
zero in any coordinate system~\cite{Gao,Zhujun}. Gauss's theorem then ensures
that the closed surface integral of the skyrmion field is zero
\begin{equation}
\label{Eq.14}
\oint \Sigma^i ds_i = 0 ,
\end{equation}
as long as there are no regions inside the surface for which the skyrmion field
is undefined. These would correspond to places for which the Stokes tensor or
its covariant derivatives are also undefined~\cite{Zhujun}.  At such places we have a 
failure of of the condition $\tensor{\Sigma}{^i_;_i} = 0$.

\begin{figure}[htbp]
\centering
\includegraphics[width=\columnwidth]{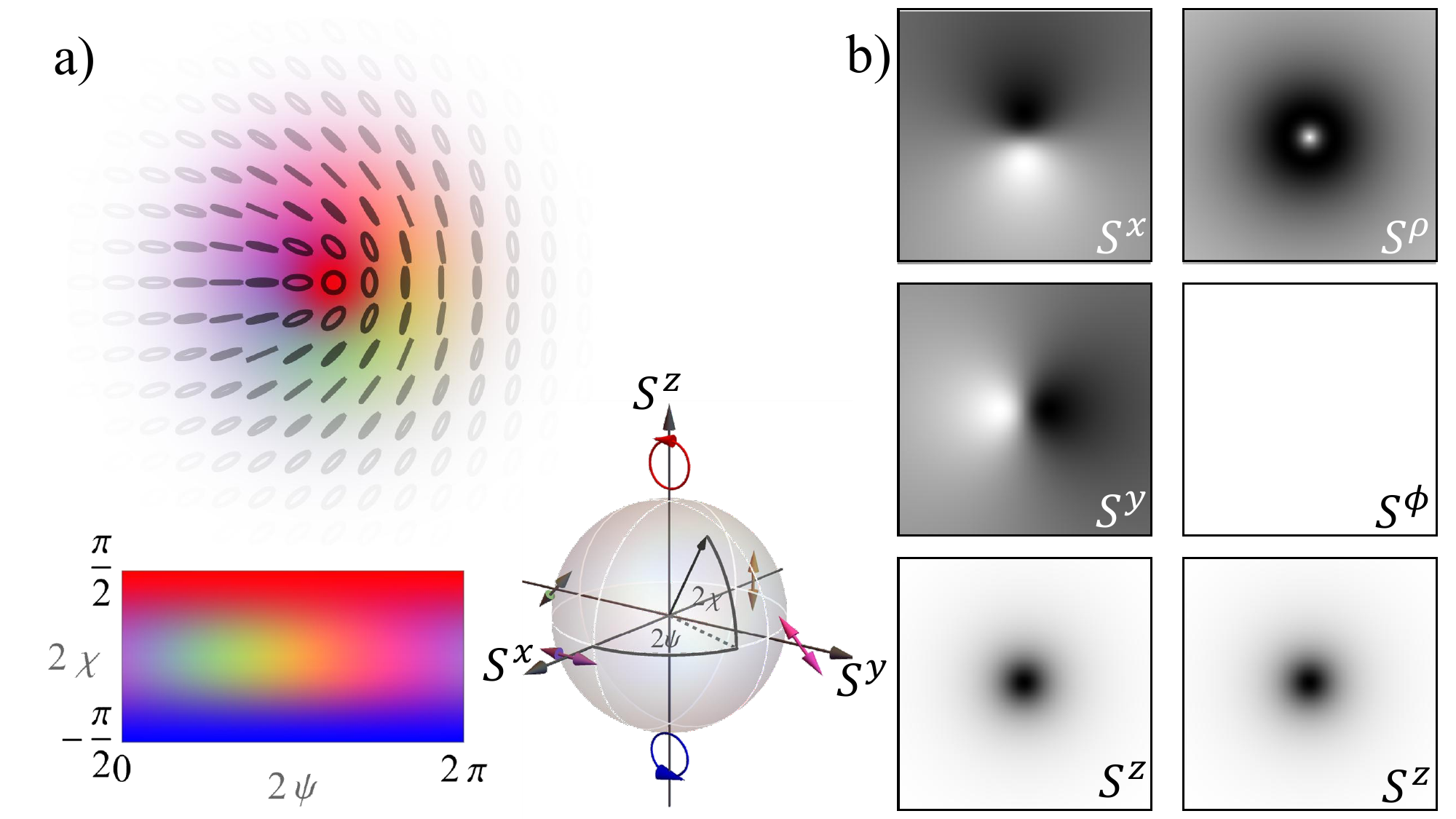}
\caption{a) Polarisation pattern for an $n=1$ skyrmion, with the intensity profile encoded as opacity, and the local polarisation as colour map. The insets show the corresponding Poincaré sphere together with its unwrapped colour map.  b) Spatial variations of the Cartesian Stokes parameters (left column), and of the corresponding Stokes tensors in cylindrical polar coordinates (right column).
}
\label{fig:figure1}
\end{figure}

The simplest paraxial skyrmion comprises two orthogonally circularly polarized
modes, one a Gaussian and the second a co-axial lowest order Laguerre-Gaussian
(LG$_{01}$) mode, the combination of which we know to have a skyrmion number of
$1$ (or $-1$)~\cite{Gao}. We can further simplify the analysis by giving the
modes the same beam waist $w_0$ and restricting our attention to the common
focal plane. With these conditions, the Cartesian components of the Stokes
tensor are~\cite{Zhujun}
\begin{align}
\label{Eq.15}
S^x &= \frac{2(\rho/w_0)\cos\phi}{1 + (\rho/w_0)^2}   \nonumber \\
S^y &= \frac{2(\rho/w_0)\sin\phi}{1 + (\rho/w_0)^2}   \nonumber \\
S^z &= \frac{1 - (\rho/w_0)^2}{1 + (\rho/w_0)^2} .
\end{align}
This displays the now familiar right circular polarization on the axis (at
$\rho = 0$) surrounded by a rotating pattern of elliptical polarization heading
towards left circular polarization as $\rho$ tends to infinity, as depicted in
Fig.~\ref{fig:figure1}. If we treat these as the Cartesian components of a
contravariant tensor, then it follows that the components of this tensor in
cylindrical polar coordinates are
\begin{align}
\label{Eq.16}
S^\rho &= \frac{2(\rho/w_0)}{1 + (\rho/w_0)^2}   \nonumber \\
S^\phi &= 0 \nonumber \\
S^z &= \frac{1 - (\rho/w_0)^2}{1 + (\rho/w_0)^2} .
\end{align}
These cylindrical-polar components are depicted in the right hand column of Fig.~\ref{fig:figure1}b), and it is clear that these are simpler
than the corresponding Cartesian components of the Stokes tensor in the left hand column.
Moreover, they display the cylindrical symmetry associated with the
skyrmion.  This is an indication of the benefit to be had with the use
of non-Cartesian coordinates and tensors.

That $S^\phi = 0$ is a consequence (in part) of the fact that the
description of the structured polarization is simpler when expressed in
cylindrical polar coordinates.  It also requires an explanation, as the form of
the skyrmion field requires all three components of the Stokes tensor to
contribute, but $S^\phi$ is zero everywhere, and there is also no remaining
$\phi$-dependence so the partial derivates with respect to $\phi$ are also
zero! This is where the difference between partial derivatives and covariant
derivatives comes in.
Working in cylindrical polar coordinates we can write the $z$-component of the
skyrmion tensor in the form
\begin{align}
\label{Eq.17}
\Sigma^z &= S^z\left(\tensor{S}{^\rho_;_\rho}\tensor{S}{^\phi_;_\phi} - \tensor{S}{^\phi_;_\rho}\tensor{S}{^\rho_;_\phi}\right)
+ S^\rho\left(\tensor{S}{^\phi_;_\rho}\tensor{S}{^z_;_\phi} - \tensor{S}{^z_;_\rho}\tensor{S}{^\phi_;_\phi}\right)  \nonumber \\
&  \qquad \qquad + S^\phi\left(\tensor{S}{^z_;_\rho}\tensor{S}{^\rho_;_\phi} - \tensor{S}{^\rho_;_\rho}\tensor{S}{^z_;_\phi}\right) .
\end{align}
To proceed, we require the non-zero components of the connection, of which there
are only two:
\begin{align}
\label{Eq.18}
\tensor{\Gamma}{^\rho_\phi_\phi} &= -\rho  \nonumber \\
\tensor{\Gamma}{^\phi_\phi_\rho} &= \frac{1}{\rho} = \tensor{\Gamma}{^\phi_\rho_\phi} .
\end{align}
It follows that the covariant derivatives are
\begin{align}
\label{Eq.19}
\tensor{A}{^i_;_z} &= \frac{\partial A^i}{\partial z}  \nonumber \\
\tensor{A}{^i_;_\rho} &= \frac{\partial A^i}{\partial \rho}  \nonumber \\
\tensor{A}{^i_;_\phi} &= \frac{\partial A^i}{\partial \phi} + \tensor{\delta}{^i_\phi} A^\rho \tensor{\Gamma}{^\phi_\phi_\rho} .
\end{align}
This means that all of the covariant derivatives in Eq.~(\ref{Eq.17}) can be
replaced with the corresponding regular partial derivatives with just one
exception:
\begin{equation}
\label{Eq.20}
\tensor{S}{^\phi_;_\phi} = \frac{\partial S^\phi}{\partial \phi} + S^\rho\tensor{\Gamma}{^\phi_\phi_\rho} = \frac{S^\rho}{\rho} .
\end{equation}
It follows that there are only two remaining, non-zero, contributions to $\Sigma^z$:
\begin{align}
\label{Eq.21}
\Sigma^z &= S^z\tensor{S}{^\rho_;_\rho}\tensor{S}{^\phi_;_\phi} - S^\rho \tensor{S}{^z_;_\rho}\tensor{S}{^\phi_;_\phi} \nonumber \\
&= \frac{S^\rho}{\rho}\left(S^z\frac{\partial S^\rho}{\partial \rho} - S^\rho\frac{\partial S^z}{\partial \rho}\right) .
\end{align}
To complete the calculation we need to evaluate only two derivatives:
\begin{align}
\label{Eq.22}
\frac{\partial S^\rho}{\partial \rho} &= \frac{2[1 - (\rho/w_0)^2]}{w_0[1 + (\rho/w_0)^2]^2}  \nonumber \\
\frac{\partial S^z}{\partial \rho} &= -\frac{4(\rho/w_0)}{w_0[1 + (\rho/w_0)^2]^2}  ,
\end{align}
so that
\begin{equation}
\label{Eq.23}
\Sigma^z = \frac{4}{w^2_0}\frac{1}{[1 + (\rho/w_0)^2]^2} .
\end{equation}
Finally, again working in cylindrical polar coordinates, the skyrmion number is
\begin{equation}
\label{Eq.24}
n = \frac{1}{4\pi}\int_0^{2\pi}\mathrm{d}\phi\int_0^\infty \rho\,\mathrm{d}\rho\,\frac{4}{w^2_0}\frac{1}{[1 + (\rho/w_0)^2]^2} = 1 ,
\end{equation}
in agreement with earlier calculations utilising Cartesian coordinates
\cite{Gao,Zhujun}.  Naturally, the application of cylindrical polar coordinates
and, indeed, other coordinate systems to determining the skyrmion tensor field,
and the associated skyrmion number, follows.

Beyond the focal plane this skyrmion number will persist, typically, as one might expect for a topological feature. It can, however, change abruptly if one of the contributing modes diffracts more strongly than the other~\cite{Gao}.

\section{Polarization beyond paraxial optics}

If we move beyond the paraxial approximation then we can typically no longer associate
a unique propagation direction with the light.  It remains the case that at any 
given point in a monochromatic field the dynamics of the electric field is 
restricted to a single plane, but the orientation of this plane varies from 
point to point \cite{Nye,NyeHajnal,BerryDennis}.  This orientation is itself 
a vector field and we can conveniently associate it with a simple cross product
of the complex electric field and its complex conjugate:
\begin{equation}
{\bf v} = -i\mbox{\boldmath$\mathcal{E}$}^*\times \mbox{\boldmath$\mathcal{E}$} .
\end{equation}
 The lines of constant linear or circular polarization are then, respectively, 
 those for which this field is zero or equal in magnitude to the modulus squared 
 of the complex electric field: $|\mathcal{E}|^2$ \cite{Nye,NyeHajnal,BerryDennis}.
We can form a vector field from the components of the normalized form of this vector:
\begin{equation}
{\bf V} = \frac{-i\mbox{\boldmath$\mathcal{E}$}^*\times \mbox{\boldmath$\mathcal{E}$}}{\sqrt{(\mbox{\boldmath$\mathcal{E}$}^*\cdot\mbox{\boldmath$\mathcal{E}$})^2
- |\mbox{\boldmath$\mathcal{E}$}\cdot\mbox{\boldmath$\mathcal{E}$}|^2}} .
\end{equation}
As with the Stokes tensor, we can treat this normalized vector field as a rank-one 
contravariant tensor and thereby obtain components in other (non-Cartesian) 
coordinate systems.  The associated skyrmion tensor is of the form in eqn.
(\ref{Eq.12}), but with the Stokes tensor replaced by $V^i$.

To illustrate this idea, let us consider one of the simplest and most fundamental 
sources of polarization structured light, an oscillating or rotating electric dipole,
which we place at the origin of our coordinate system.  The complex electric and
magnetic fields for this source have the form \cite{Jackson}
\begin{align}
\label{Eq.25}
\mbox{\boldmath$\mathcal{E}$}({\bf r})  &= \frac{e^{-i\omega(t-r/c)}}{4\pi\varepsilon_0r^3}
\Bigg[({\bf d} - 3({\bf d}\cdot\hat{\bf r})\hat{\bf r})\left(1 - \frac{i\omega r}{c}\right) \nonumber \\
& \quad \quad \quad \quad \quad \quad \quad \quad - ({\bf d} - ({\bf d}\cdot\hat{\bf r})\hat{\bf r})\left(\frac{\omega r}{c}\right)^2\Bigg]  \nonumber \\
\mbox{\boldmath$\mathcal{H}$}({\bf r})  &=  \frac{i\omega e^{-i\omega(t-r/c)}}{4\pi r^2}
(\hat{\bf r}\times{\bf d})\left(1 - \frac{i\omega r}{c}\right) ,
\end{align}
where ${\bf d}$ is the (in general complex) dipole moment and $\omega$ is the angular
oscillation frequency of the dipole.  A straightforward calculation shows that in
this case we have
\begin{equation}
-i\mbox{\boldmath$\mathcal{E}$}^*\times \mbox{\boldmath$\mathcal{E}$}
= \frac{-i\omega^4}{(4\pi\varepsilon_0c^2r)^2}[{\bf d}^*\times{\bf d}
- (\hat{\bf r}\cdot{\bf d}^*)\hat{\bf r}\times{\bf d}
+ (\hat{\bf r}\cdot{\bf d})\hat{\bf r}\times{\bf d}^*]  .
\end{equation}
This is clearly zero if the dipole moment, ${\bf d}$, is real.  For a rotating dipole 
it will have a non-zero value.  To illustrate this point, let the dipole be 
rotating in the $x-y$ plane so that
\begin{equation}
\label{Eqdipole}
    {\bf d} = \frac{d}{\sqrt{2}}(\hat{\bf x} + i\hat{\bf y}), \qquad 
    {\bf d}^* = \frac{d}{\sqrt{2}}(\hat{\bf x} - i\hat{\bf y}).
\end{equation}
In this case we find that our field is purely radial, with the form
\begin{equation}
    -i\mbox{\boldmath$\mathcal{E}$}^*\times \mbox{\boldmath$\mathcal{E}$}
    = \frac{\omega^4d^2}{(4\pi \varepsilon_0c^2r)^2}\cdot\frac{z}{r} \,
    \hat{\bf r} .
\end{equation}
so that the corresponding normalized form is
\begin{equation}
    {\bf V} = {\rm sign}(z)\hat{\bf r} .
\end{equation}
When considered as a rank-one contravariant tensor, this has the components, in 
spherical polar coordinates:
\begin{equation}
    V^r = {\rm sign}(z),  \quad V^\theta = 0, \quad V^\phi = 0 .
\end{equation}
It follows that the Skyrmion tensor has only a radial component, which is 
well-behaved everwhere except in the $x-y$ plane:
\begin{align}
\Sigma^r &= V^r(V^\theta_{;\theta}V^\phi_{;\phi} - V^\theta_{;\phi}V^\phi_{;\theta})
\nonumber \\
&= V^r(V^r\Gamma^\theta_{r\theta}V^r\Gamma^\phi_{r\phi} 
- V^r\Gamma^\theta_{r\phi}V^r\Gamma^\phi_{r\theta})  \nonumber \\
&= \frac{{\rm sign}(z)}{r^2} ,
\end{align}
where we have used the elements of the connection for spherical polar
coordinates: $\tensor{\Gamma}{^\theta_r_\theta} = r^{-1} =
\tensor{\Gamma}{^\phi_r_\phi}$ and $\tensor{\Gamma}{^\theta_r_\phi} = 0 =
\tensor{\Gamma}{^\phi_r_\theta}$.  It follows that the skyrmion number based
on integration over a full sphere enclosing and centred on the dipole has 
the value
\begin{equation}
    n = \frac{1}{4\pi}\oint \Sigma^i ds_i = 0 ;
\end{equation}
Yet there is a natural structure to be found in the polarization field that is
associated with the flux of the skyrmion field, separately, through the 
upper (positive $z$) and lower (negative $z$) half spheres.  In particular the 
flux of the Skyrmion field through the upper hemisphere is
\begin{equation}
    \frac{1}{4\pi}\int_{z\geq0}\Sigma^ids_i = \frac{r^2}{4\pi}\int_0^{2\pi}d\phi
    \int_0^{\pi/2}\sin\theta d\theta \: \Sigma^r = \frac{1}{2} ,
\end{equation}
and the corresonding value for the lower half sphere is $-1/2$.  This makes good
physical sense, for the vector field ${\bf v}$ is both related to the optical spin 
density and also the helicity flux \cite{Alison,Frances}.  The rotating dipole 
can and does radiate optical angular momentum \cite{OAMFlux}, but in opposite 
directions for the upper and lower half spheres.  It is these opposing directions 
that cause the two skyrmion fields, which derive from the helicity fluxes, to
cancel each other.

\section{Skyrmions beyond polarization}

From a mathematical perspective, the Skyrmion formalism can be applied to any vector 
field, with the elements corresponding to the Stokes parameters being, simply, the 
direction in which the vector is orientated at any given point in space.  To 
demonstrate this idea we consider two elementary examples from distinct areas
of physics: the Poynting vector associated with the radiating dipole considered 
above, and the Newtonian gravitatonal field associated with a point mass.

\subsection{Skyrmion field based on Poynting's vector} 
\label{sec_5} 

The use of the skyrmion tensor is not restricted to phenomena associated with
polarization.  As an example of this versatility, we construct a skyrmion field 
for Poynting's vector for our electric dipole source. We can express Poynting's vector
in the form
\begin{equation}
    {\bf P} = \Re\left(\mbox{\boldmath$\mathcal{E}$}^*\times\mbox{\boldmath$\mathcal{H}$}\right) .
\end{equation}
(We use ${\bf P}$ for Poynting's vector in
place of the more usual ${\bf S}$ to avoid any possible confusion with the
Stokes vector).  For our electric dipole source this takes the form
\begin{equation}
\begin{split}
{\bf P} 
= \frac{\omega}{16\pi^2\varepsilon_0r^5}
\Bigg\{ & \hat{\bf r}(d^2 - |{\bf d}\cdot\hat{\bf r}|^2)\left(\frac{\omega r}{c}\right)^3 \\
& + i[{\bf d}^*({\bf d}\cdot\hat{\bf r}) - {\bf d}({\bf d}^*\cdot\hat{\bf r})]
\left(1 + \left(\frac{\omega r}{c}\right)^2\right)\Bigg\} .
\end{split}
\end{equation}
For a purely real dipole moment, the second term is zero and we find that the 
Poynting vector is purely radial, corresponding to radial emission of light
from the centrally-placed dipole.  If the dipole moment is complex, however, then
we have a second, mutually orthogonal component.  To make physical sense of this,
let us return to the rotating dipole in the $x-y$ plane with a complex dipole 
vector of the form in eqn.~(\ref{Eqdipole}).  In this case there are two non-vanishing
components of the Poynting tensor in spherical polar coordinates:
\begin{align}
    P^r &= \frac{\omega^4 d^2}{16\pi^2\varepsilon_0c^3r^2}
    \left(1 - \frac{1}{2}\sin^2\theta\right)  \nonumber \\
    P^\phi &= \frac{\omega d^2}{32\pi^2\varepsilon_0c^3r^6}
    \left[1 + \left(\frac{\omega r}{c}\right)^2\right] .
\end{align}
The existence, for a rotating dipole, of an azimuthal component to Poynting's 
vector is intimately connected with the fact that, in this case, the radiated
light carries both linear and orbital angular momentum.

In the far field, $(r \gg c/\omega)$, the Poynting vector retains only the single component in the radial direction and it follows that the normalized components are
then simply 
\begin{equation}
\label{Eq.27a}
\hat{P}^r = 1 , \quad \hat{P}^\theta = 0 , \quad \hat{P}^\phi = 0 .
\end{equation}
It follows that the Poynting-tensor based skyrmion field tensor has the single non-zero component
\begin{align}
\label{Eq.28}
\Sigma^r &= \hat{P}^r\left(\tensor{\hat{P}}{^\theta_;_\theta}\tensor{\hat{P}}{^\phi_;_\phi} - \tensor{\hat{P}}{^\theta_;_\phi}\tensor{\hat{P}}{^\phi_;_\theta}\right)
\nonumber \\
&= \hat{P}^r\left(\hat{P}^r\tensor{\Gamma}{^\theta_r_\theta}\hat{P}^r\tensor{\Gamma}{^\phi_r_\phi} -
\hat{P}^r\tensor{\Gamma}{^\theta_r_\phi}\hat{P}^r\tensor{\Gamma}{^\phi_r_\phi}\right)  \nonumber \\
&= \frac{1}{r^2} ,
\end{align}
It then follows that the skyrmion number associated with the Poynting tensor for an oscillating dipole is unity:
\begin{align}
\label{Eq.29}
n &= \frac{1}{4\pi}\oint \Sigma^ids_i \nonumber \\
&= \frac{r^2}{4\pi} \int_0^{2\pi} \mathrm{d}\phi\int_0^\pi \sin\theta\,\mathrm{d}\theta  \: \Sigma^r  \nonumber \\
&= 1 ,
\end{align}
where we have exploited the spherical symmetry of the problem by integration over a spherical shell
centred on the dipole.

In the near field, the Poynting tensor acquires a non-zero $\phi$ component. This is responsible for the existence of angular momentum in the radiated field~\cite{Les}. It will also modify the direction of the corresponding Skyrmion field.

It is interesting to note that although
$\mbox{\boldmath$\nabla$}\cdot\mbox{\boldmath$\Sigma$}=0$, we have a non-zero
closed-surface integral of $\mbox{\boldmath{$\Sigma$}}$, which seems like a
violation of Gauss's law.  That this is not the case follows from the fact that
the direction of the Poynting vector is not defined at the origin and that
therefore its derivatives are similarly not defined.  At such points 
we have a failure of the condition 
$\mbox{\boldmath$\nabla$}\cdot\mbox{\boldmath$\Sigma$}=0$ and the origin acts 
as a source of the skyrmion field.  This is analogous to the
application of the first Maxwell equation to determine the electric field of a
point charge, where $\mbox{\boldmath$\nabla$}\cdot{\bf E}=0$ fails at the position
of the charge.  The magnetism community refer to this singular point as a Bloch point
\cite{Yasin}.  We note that the same issue arises when treating paraxial
optical skyrmions with non-integer skyrmion number, for which the polarization,
and hence the Stokes vector, is not defined on the beam axis~\cite{Zhujun}.

\subsection{Skyrmion field for Newtonian gravity}

We have seen that the naturally outward radial form of Poynting's vector for a 
dipole source corresponds to a radially outwards skyrmion tensor.  With this 
comes an associated skyrmion number of one, associated with any surface enclosing the 
dipole.  A similar radial tensor can be associated with the gravitational field of a 
point mass, $m$, placed at the origin.  For this situation we have the gravitational 
field
\begin{equation}
{\bf G} = -\frac{Gm}{r^2}\hat{\bf r} .
\end{equation}
It follows that the normalized components of this field considered as a tensor
are
\begin{equation}
    \hat{G}^r = -1 \quad \hat{G}^\theta = 0 \quad \hat{G}^\phi = 0 .
\end{equation}
The corresponding skyrmion tensor has the single non-zero component
\begin{equation}
    \Sigma^r = -\frac{1}{r^2} ,
\end{equation}
so the skyrmion flux through a surface enclosing the mass is
\begin{equation}
    n = \frac{r^2}{4\pi}\int_0^{2\pi}d\phi\int_0^\pi \sin\theta d\theta \: \Sigma^r = -1.
\end{equation}
This non-zero value is associated with the point at the origin where the 
skyrmion field and, indeed, the direction of the gravitational field, is
undefined. This is not a problem, physically, as at this point the gravitational field is zero. In the magnetism community, this singular point is referred to
as an anti-Bloch point \cite{Yasin}.

\section{Conclusion}
The introduction of tensors and the associated covariant derivatives makes it
both possible and convenient to work with non-Cartesian coordinates, especially
when they reflect the symmetry of the physical system. We have seen that this
allows us to describe the Stokes parameters and associated skyrmion fields for
simple patterns of structured light in paraxial optics by utilizing cylindrical
polar coordinates. This should also simplify the study of the propagation of
such light beams. 

The application of tensorial methods is not restricted to paraxial optics.
We have illustrated this by presenting the form of the skyrmion tensor for
the cross product of the complex electric field with its complex conjugate.
For the light emitted by a rotating dipole this quantity is simply related to
the helicity flux in the radiated field \cite{Frances}.  An oscillating 
electric dipole also radiates energy, with the energy flux being given by
Poynting's vector.  This vector field has, also, an associated skyrmion
tensor that is simply related to the direction of the outgoing radiation.

It seems clear that skyrmions are a usual characteristic feature of structured
light fields, but that they do not always appear in quantities that have so far
been investigated. By freeing the theory of optical skyrmions from the
constraints of Cartesian coordinates, we may hope that the ideas embodied in
the theory of skyrmions may find broader application in optics, for example structures formed from Bessel beams~\cite{BesselSkyrmions} and accelerating Airy beams~\cite{AiryBeams}, among others~\cite{ParticleLike}.

Finally, we note that skyrmions have found wide application outside the
field of optics, and that the use of our tensor formulation might fruitfully 
find application there.  We note, in particular the possibility of application 
in the field of magnetic skyrmions, with the Stokes tensor replaced by
a unit magnetisation tensor \cite{Sachdev,Seki}.  As with optical skyrmions,
the tensorial formulation is well adapted to situations with cylindrical
or spherical symmetry \cite{Berganza}.

\begin{backmatter}
\bmsection{Funding}
This work was supported, in part, by the COST Action POLYTOPO: CA23134, Topological textures 
in condensed matter.
SMB thanks the Royal Society for support RSRP/R/210005. SF-A acknowledges support through the QuantERA II Programme, with funding received via the EU
H2020 research and innovation programme under grant 101017733, and associated
support from EPSRC under grant EP/Z000513/1 (V-MAG). FCS acknowledges support from EPSRC under grant EP/Y014456/1.
\bmsection{Acknowledgments}  We thank Markus Garst and Jan Masell for helpful
discussions and suggestions.
\bmsection{Disclosures} The authors declare no conflicts of interest.

\bmsection{Data Availability Statement}  No data were generated or analyzed in
the presented research.
\end{backmatter}


\end{document}